\documentclass[iop,apjl]{emulateapj}
\usepackage{ulem}

\newcommand{\simless}{\mathbin{\lower 3pt\hbox
      {$\rlap{\raise 5pt\hbox{$\char'074$}}\mathchar"7218$}}} %< or of order
\newcommand{\simgreat}{\mathbin{\lower 3pt\hbox
     {$\rlap{\raise 5pt\hbox{$\char'076$}}\mathchar"7218$}}} %> or of order

\slugcomment{\today}

\shorttitle{ALMA observations of $\rho$-Oph~102}
\shortauthors{Ricci et al.}

\begin{document}


\title{ALMA observations of $\rho$-Oph~102: grain growth and molecular gas in the disk around a young Brown Dwarf}


\author{L. Ricci\altaffilmark{1}, 
L. Testi\altaffilmark{2,3}, 
A. Natta\altaffilmark{3,4},
A. Scholz\altaffilmark{4} and
I. de Gregorio-Monsalvo\altaffilmark{5}}

\altaffiltext{1}{Department of Astronomy, California Institute of Technology, MC 249-17, Pasadena, CA 91125, USA}
\altaffiltext{2}{European Southern Observatory, Karl-Schwarzschild-Strasse 2, D-85748 Garching, Germany}
\altaffiltext{3}{INAF-Osservatorio Astrofisico di Arcetri, Largo E. Fermi 5, I-50125 Firenze, Italy}
\altaffiltext{4}{School of Cosmic Physics, Dublin Institute for Advanced Studies, 31 Fitzwilliam Place, Dublin 2, Ireland}
\altaffiltext{5}{Joint ALMA Observatory (JAO)/ESO. Alonso de Cordova 3107. Vitacura 763 0335. Santiago de Chile}
%\altaffiltext{5}{European Southern Observatory, Vitacura Casilla 19001, Santiago de Chile 19, Chile}
\email{lricci@astro.caltech.edu}


\begin{abstract}

\noindent We present ALMA continuum and spectral line observations of the young Brown Dwarf $\rho$-Oph~102 at about 0.89~mm and 3.2~mm. We detect dust emission from the disk at these wavelengths and derive an upper limit on the radius of the dusty disk of $\sim 40$~AU.
The derived variation of  the dust opacity with frequency in the mm provides evidence for the presence of mm-sized grains in the disk outer regions. This result demonstrates that mm-grains are found even in the low density environments of Brown Dwarf disks and challenges our current understanding of dust evolution in disks.
The CO map at 345 GHz clearly reveals molecular gas emission at the location of the Brown Dwarf, indicating a gas-rich disk as typically found for disks surrounding young pre-Main Sequence stars. 
We derive a disk mass of $\sim 0.3-1$\%\ of the mass of the central Brown Dwarf, similar to the typical values found 
for disks around more massive young stars.

\end{abstract}


\section{Introduction}
\label{sec:intro}

Brown Dwarfs (BDs) are very low-mass (M$\le$0.085M$_\odot$), stellar-like objects unable to burn hydrogen.
%, some of them are below the deuterium burning limit and have masses comparable with those of giant planets. 
%Being at the low boundary of the stellar mass spectrum, free floating young BDs are an ideal laboratory to study an extreme case of the formation and early evolution of stars. In low-mass stars, circumstellar disks are formed as a consequence of the conservation of angular momentum from the collapsing parental core. In the first few million years of evolution of the central star, these disks contain $\sim 0.1 - 20$\%\ of the central object mass \citep{WilliamsCieza:2011}. Planetary systems are believed to form out of this reservoir of material. The picture for BDs is more uncertain on both their origin and their ability of form planetary systems. 
It is now established that young BDs are surrounded by dusty circumstellar disks \citep{NattaTesti:2001,Natta:2002,Testi:2002,Scholz:2008} and undergo a T Tauri-like phase during their early evolution that involves disk accretion and outflows \citep{Jayawardhana:2003,Natta:2004}. 
%Nevertheless, while it is likely that there is a continuity of properties from low-mass stars through the BDs regime, it is  unclear whether the majority of young BDs form like stars from isolated cores or 
%whether a significant fraction of the BD population is produced when accretion on a stellar embryo is abruptly terminated, as in the 
%case of ejection from a disk or a multiple (proto-)stellar system 
%through an ejection from either a multiple system or a disk \citep{ReipurthClarke:2001,Stamatellos:2007,Bate:2009}. 
%Ejection models predict young BDs disks populations with small radii ($\le$10~AU) and small masses. 

Because of their very low masses, disks around BDs also represent an extreme environment to test planet formation theories. There are a few wide orbit, giant planets known as companions to BDs, most likely formed as a result of the fragmentation of the proto-BD core or by gravitational instabilities in the outer disk \citep[e.g. 2M1207b, 2M0441b,][]{Chauvin:2005,Lodato:2005,Todorov:2010}. However, most extrasolar planets (and all rocky planets) are thought to form via the core accretion process \citep{Matsuo:2007}, and evidence of the first steps of this process has been found in disks around more massive pre-Main Sequence (PMS) stars. In particular, sub-mm observations, which probe the bulk of the disk cold material,  have shown that the grains have grown to  to reach millimeter and centimeter size \citep[e.g.][]{Beckwith:1991,Rodmann:2006,Ricci:2010a,Ubach:2012}. 

In BD disks, infrared spectroscopy has shown that in several sources silicates on the disk surface have been significantly modified in size and crystallinity, as in PMS stars \citep{Sterzik:2004,Apai:2005}. Until now, due to the limited sensitivity of sub-mm telescopes, the detection of long wavelength emission from young BDs has been limited to continuum photometry of the few brightest objects \citep[e.g.][]{Scholz:2006,Mohanty:2012}. The start of ALMA operations allows us to perform more detailed studies. Here we report on multi-wavelength continuum and CO($J=3-2$) observations on the young BD $\rho$-Oph~102. This object ($\sim 60$~M$_{Jup}$, M6-spectral type) in the $\rho$-Ophiuchi star forming region (SFR) is known to be surrounded by a disk from infrared observations, to have a significant mass accretion rate and to drive a wind and molecular outflow \citep[][]{Bontemps:2001,Natta:2002,Natta:2004,Whelan:2005,Phan-Bao:2008}. In this Letter we report on our ALMA observations, which were designed to derive a solid estimate of the mm spectral index to constrain the dust properties. In particular we found evidence for grain growth to mm-grain sizes in the outer regions of the $\rho$--Oph 102 disk. 
Also, we clearly detected molecular gas emission from the disk and derived an estimate for its mass.

% to constrain the gas content of the disk. 
%The ALMA observations and results are presented in Sect.~\ref{sec:obs}, and are then then discussed in Sect.~\ref{sec:discussion}.

\noindent 
\section{ALMA Observations and Results}
\label{sec:obs}

\subsection{Observations and data reduction}

We observed $\rho$--Oph~102 using ALMA Early Science in Cycle~0 at Band~7 and~3 (about 345 and 100~GHz, respectively). Observations in Band~7 were performed using 15 antennas in the Compact array configuration on November 3, 2011, and 16 antennas in the Extended array configuration on May 23, 2012 (projected baseline lengths in the range from the shadowing limit to $\sim$400~m). 
Observations in Band~3 were conducted with 16 antennas in the Extended configuration on May 8, 2012 (projected baseline lengths from  $\sim$35 to $\sim$402~m). All observations were done in good and stable weather conditions, with precipitable water vapor of $\sim 0.8-1.2$ mm and $\sim 1.9$ mm at Band ~7 and 3, respectively.
The ALMA correlator was set to record dual polarization with four separate spectral windows, each providing a bandwidth of 1.875~GHz with channels of 0.488~MHz width. Spectral windows were centered at 331.103, 332.998, 343.103, 344.998 GHz for Band~7 observations and 86.1021, 87.9976, 98.103, 99.999 GHz for Band~3.
The total integration times on $\rho$--Oph~102 were approximately 15 and 30 minutes in Band~7 and~3, respectively. 

The interferometric visibility data were reduced using the CASA package \citep[][]{McMullin:2007}. J1924-292 and 3C279 were observed as bandpass calibrators, Neptune and Titan for flux calibration. Simultaneous observations of the 183~GHz water line with the water vapour radiometers were used to reduce atmospheric phase noise before using J1625-254 for standard complex gains calibration. The  flux scale was tied to the Butler-JPL-Horizons 2010 models of Neptune and Titan, resulting in an accuracy of  $\sim10\%$. 

Imaging of the calibrated visibilities was done in CASA. For each band, a continuum map was produced using natural weighting and combining all the channels without line emission. The only line emission detected was the CO ($J=3-2$) rotational line in Band~7 which was imaged separately. The CO(3-2) shows significant extended emission as well as compact structure. We imaged the CO data in two separate ways to highlight the different components: we first used all the data, natural weighting and a gaussian taper of the visibilities to recover as much as possible of the extended emission and compare with the low angular resolution map of \citet{Phan-Bao:2008}; additionally, we also imaged the data from baselines exceeding $\sim60$~k$\lambda$ 
%($\sim$180~m) 
with robust weighting to filter out emission on scales larger than about $3''$, or $\sim 400$ AU at the Ophiuchus distance, and highlight the compact emission associated with $\rho$--Oph~102.

\noindent

%\section{Results}
%\label{sec:res}

\subsection{Continuum maps}
\label{sec:cont}

\begin{figure}
%\plotone{band3}
\epsscale{1.1}
\plottwo{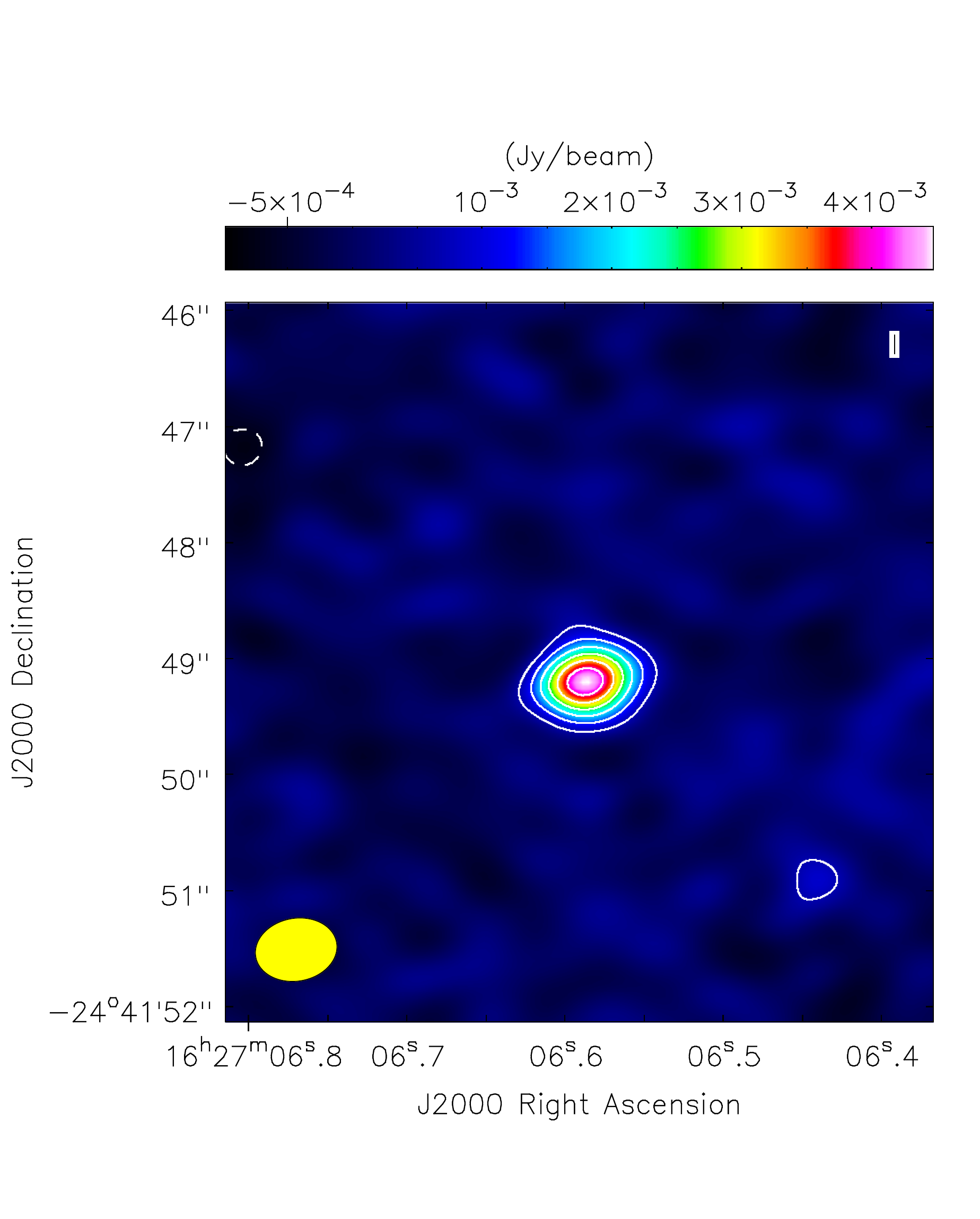}{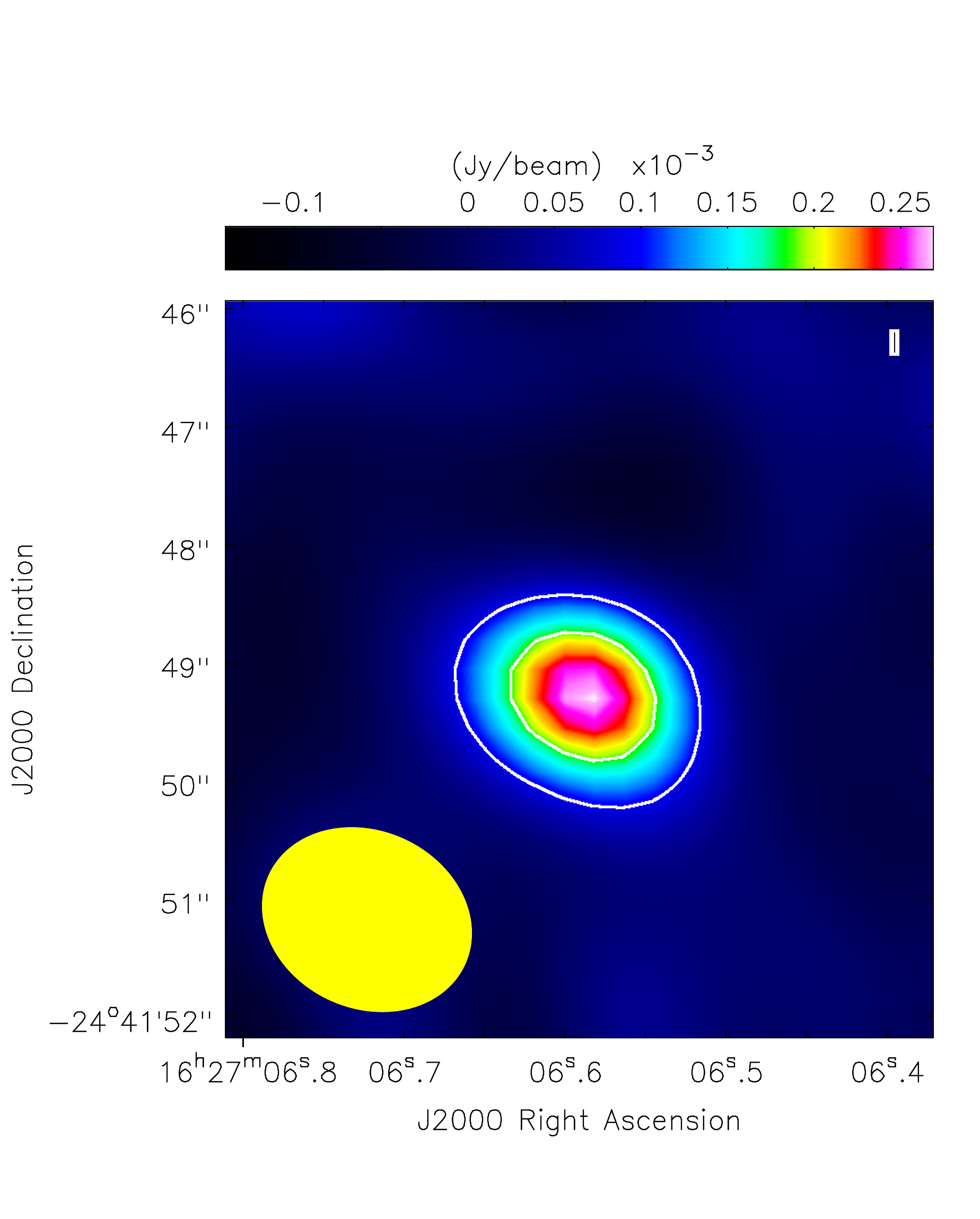}
\caption{Continuum maps of $\rho$-Oph~102. Left panel) Continuum map at 0.89~mm. White contour lines are drawn at $-3, 3, 6, 9, ..., 18\sigma$, where $\sigma = 0.22$~mJy/beam is the rms noise measured on the map.  Right) Continuum map at 3.2~mm. White contour lines are drawn at 3 and 6$\sigma$, where $\sigma = 0.031$~mJy/beam. In each panel, the yellow filled ellipse in the lower left corner indicates the size of the synthesized beam, i.e. FWHM $= 0.71'' \times 0.54''$, PA $= 100$~deg at 0.89~mm, and FWHM $=1.82'' \times 1.50''$, PA $= 66$~deg at 3.2~mm. For both maps, a Briggs weighting with robust parameter $= 2$ (natural weighting) was used to maximize the signal-to-noise ratio.}
\label{fig:cont}
\end{figure}

\begin{figure}
%\plotone{band3}
\epsscale{1.0}
\plotone{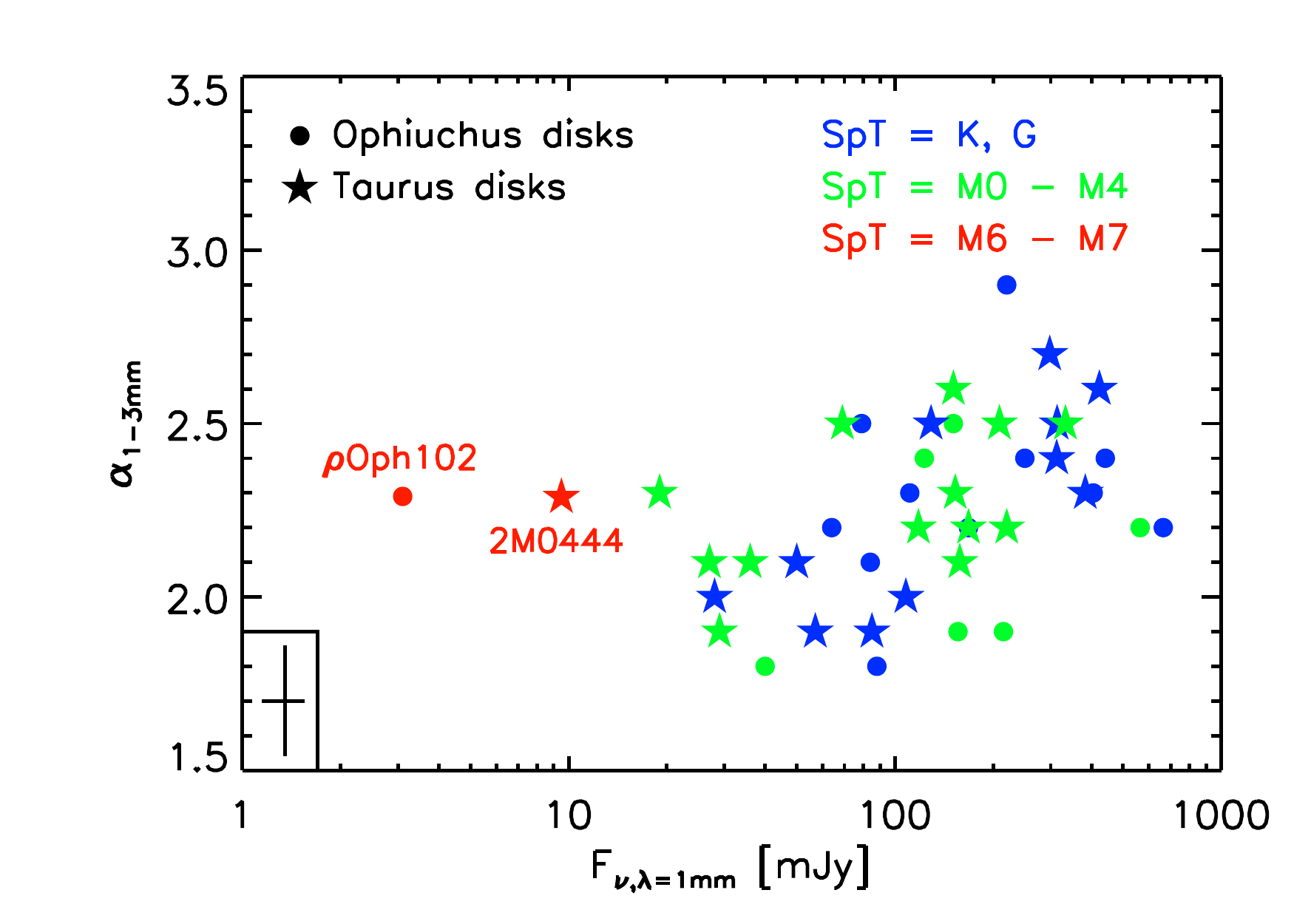}
\caption{Flux at 1 mm vs spectral index between 1 and 3 mm for disks around single PMS stars and Brown Dwarfs. Different colors and symbols refer to different stellar/sub-stellar spectral types and regions as indicated in the plot. Data for Taurus disks are from \citet{Ricci:2010a,Ricci:2012}, for Ophiuchus disks from \citet{Ricci:2010b}, for 2M0444 from \citet{Bouy:2008} and Ricci et al. (2012, in prep.), and for $\rho$-Oph 102 from this Letter. Note that for same disks the values of the 1 mm flux density has been derived by interpolating between nearby wavelengths. The typical uncertainties of the data are shown in the lower left corner.}
\label{fig:flux_alpha}
\end{figure}

Figure~\ref{fig:cont} shows the continuum maps at about 0.89~mm (Band 7) and 3.2~mm (Band 3). Dust emission from the disk surrounding $\rho$-Oph~102 is clearly detected in both maps. 
We measured flux densities of $4.10 \pm 0.22$~mJy at 0.89~mm and of $0.22 \pm 0.03$~mJy at 3.2~mm. The obtained rms-noise levels are consistent with the theoretical expectations given the on-source integration times. 
This gives a sub-mm spectral index $\alpha$ ($F_{\nu} \propto \nu^{\alpha}$) between 0.89 and 3.2~mm of $\alpha_{\rm{0.89-3.2mm}} = 2.29 \pm 0.16$, where the uncertainty accounts for both the flux calibration and the rms noise in the maps. This value lies well within the range of values measured for disks around single PMS stars in the $\rho$-Oph and Taurus SFRs (see Figure \ref{fig:flux_alpha}). This suggests that the physical properties of the dust grains emitting in the sub-mm, e.g. their size, do not significantly vary in disks around PMS stars and BDs. The implications of this result will be discussed in Section~\ref{sec:discussion}.

Using observations with the Submillimeter array \citep[SMA,][]{Ho:2004} \citet[][]{Phan-Bao:2008} reported a low signal-to-noise flux density of $7 \pm 3$~mJy at about 1.3~mm. Interpolating between our two measurements, we expect a flux density of $1.7 \pm 0.3$ mJy at 1.3~mm, lower than the value previously reported, although the discrepancy is only at the 2$\sigma$-level. Note that already with $\sim 15$ minutes on-source with ALMA in Early Science we reached a sensitivity $\sim 10$ times better than the SMA.
%Because of this, the value of the disk mass inferred in this paper is significantly lower than what derived by \citet[][see Section 4]{Phan-Bao:2008}.

By inspecting the visibility data, within the uncertainties, we did not find evidence for a decrease of the visibility amplitude with projected baseline length, neither at the shorter nor at the longer baseline lengths probed by our observations. We conclude that our observations did not spatially resolve the disk continuum emission. Considering our angular resolution of $\sim 0.6''$ at 0.89~mm, and a distance of 130~pc for the $\rho$-Oph SFR \citep[][]{Wilking:2008,Lombardi:2008}, this indicates that the dust is concentrated within less than $\sim 40$~AU from the central BD. %If confirmed with observations at higher sensitivity and angular resolution, 
%This suggests that the disk around $\rho$--Oph~102 is significantly smaller than disks normally found around single more massive PMS stars \citep[see e.g.][]{Isella:2009,Andrews:2010,Guilloteau:2011}.  

\subsection{CO molecular gas map}
\label{sec:co}

%We present the detection of the $J = 3 - 2$ rotational transition line from CO molecular gas in the surroundings of the $\rho$--Oph~102 brown dwarf. As discussed in Sect.~\ref{sec:obs} we produced images using two separate weighting schemes and baseline ranges to attempt to recover as much of the extended emission as possible and to highlight the compact emission from the material around the young brown dwarf. 

The low angular resolution, tapered natural weighting map shows a morphology of the CO($J=3-2$) that closely resembles the map of the  CO($J=2-1$) emission presented by \citet[][]{Phan-Bao:2008}.  We confirm the same main emission features as observed in the SMA map. They interpreted these structures as a bipolar molecular outflow emitted by $\rho$ Oph 102. Even if the signal to noise in our map is significantly better than in the SMA data, our data are most likely significantly affected by missing flux at short baselines. Furthermore, the most prominent features attributed to the outflow are close to the edge of the ALMA primary beam, making it difficult to accurately determine the emission and its extent. For these reasons, we will not discuss the extended emission and outflow further in this Letter.
%Figure~\ref{fig:co_map} shows the CO($J=3-2$) emission inside the primary beam of the ALMA observations. In the map, the imaging was obtained through a weighting scheme which highlights the diffuse emission found around the brown dwarf. The CO emission is dominated by two extended structures at north-east and south of $\rho$ Oph 102. The morphology, location and radial velocities of the northern and southern lobe structures closely resembles the morphology of the CO($J=2-1$) emission observed by \citet[][]{Phan-Bao:2008}. They interpreted these structures as a bipolar molecular outflow emitted by $\rho$ Oph 102. 
%However, since these lobes are located very close to the primary beam edge of our observations, it is likely that our observations have missed a significant fraction of the emission. For this reason we do not discuss these structures any further in this paper.

\begin{figure}
%\plotone{band3}
\epsscale{1.0}
\plottwo{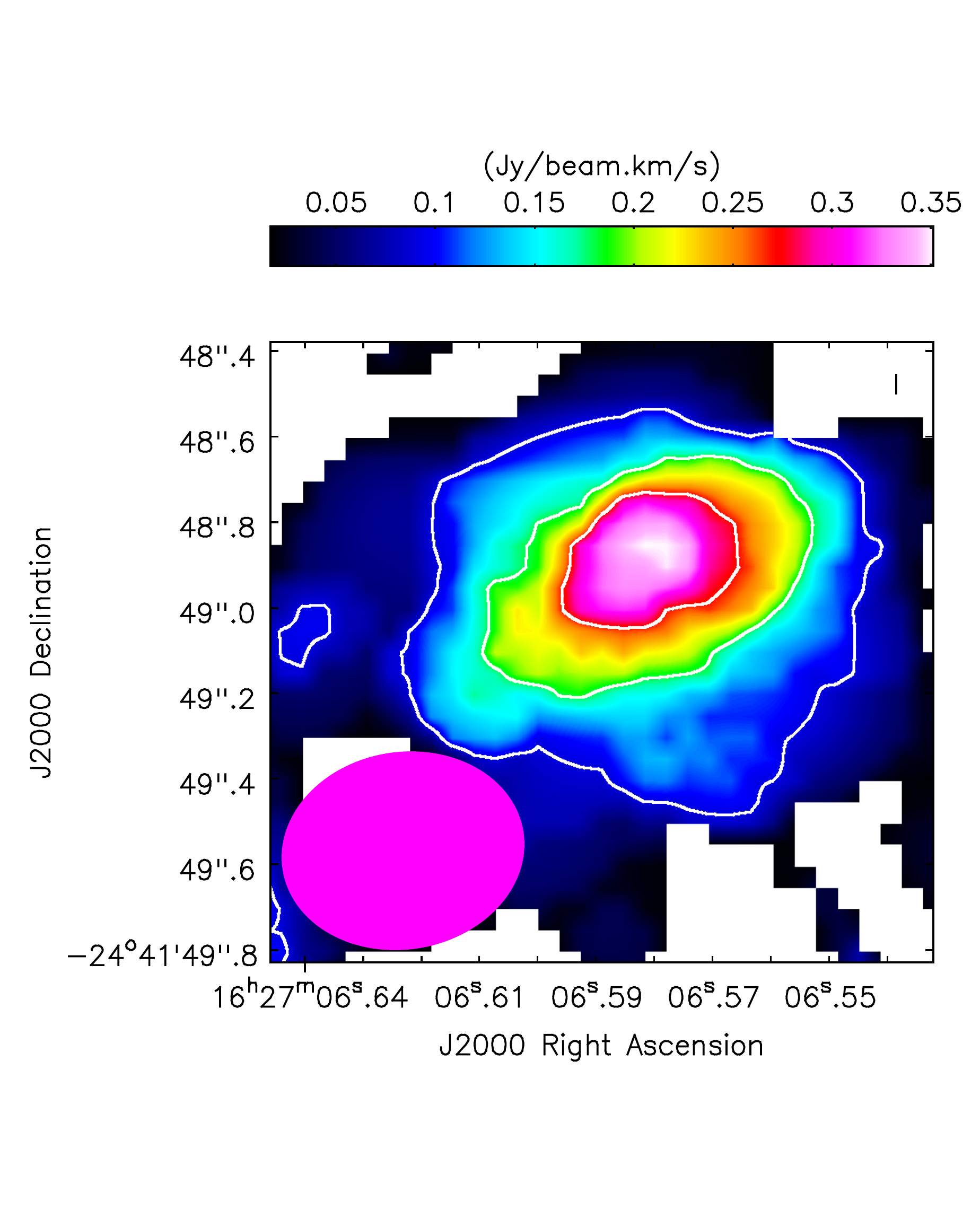}{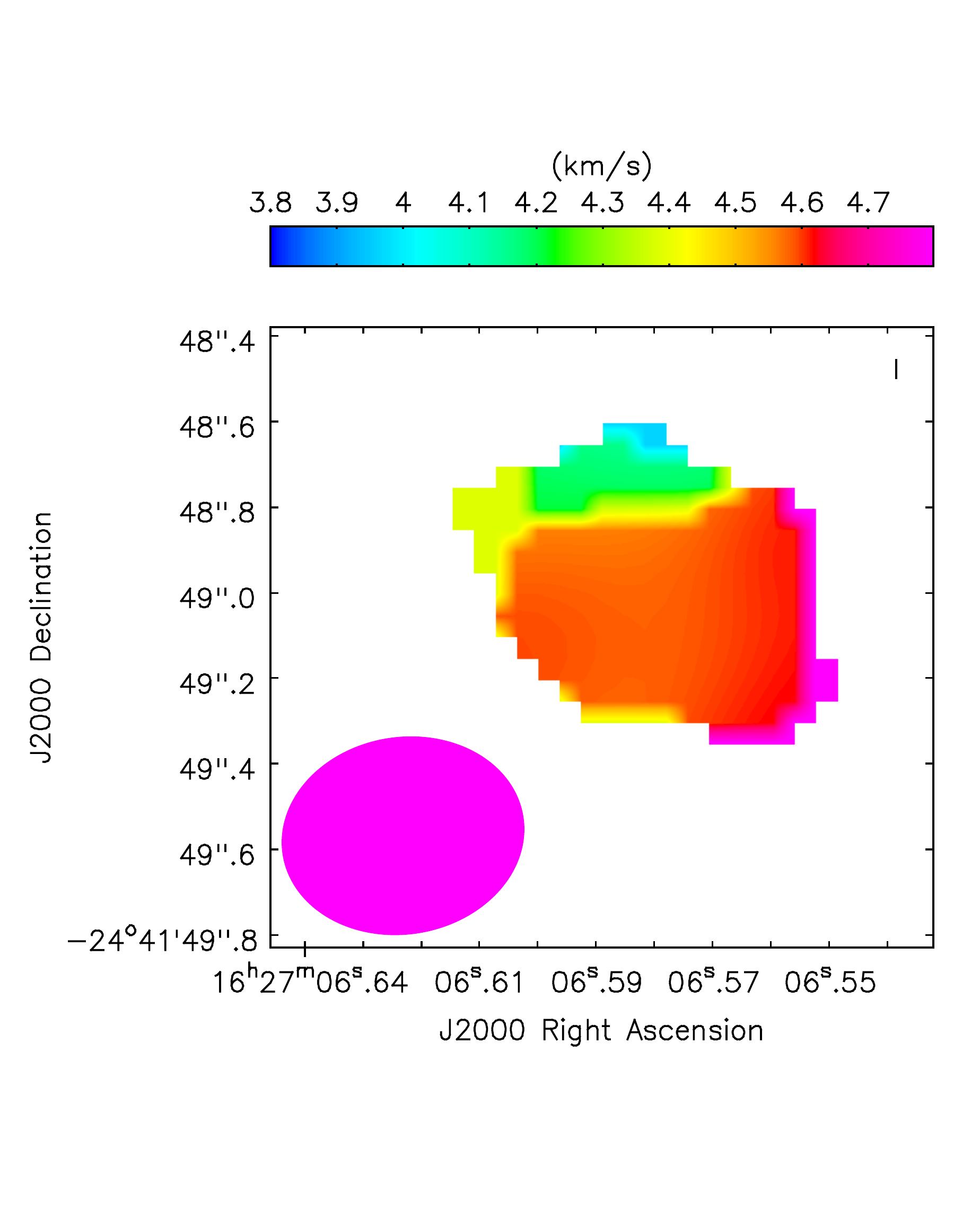}
\caption{CO($J=3-2$) maps of $\rho$-Oph~102. Left panel) Moment 0 CO($J=3-2$) map. White contour lines are drawn at $2, 4, 6\sigma$, where $\sigma = 45$~mJy/beam$\cdot$km/s is the measured rms-noise.  Right) Moment 1 CO($J=3-2$) map. In each panel, the magenta filled ellipse in the lower left corner indicates the size of the synthesized beam, i.e. FWHM $= 0.57'' \times 0.46''$, PA $= 99$~deg. For the imaging, we adopted a Briggs weighting with robust parameter $= 0$ and considered only projected baselines longer than 70~k$\lambda$ to highlight the emission from structures with angular scales $\simless~3''$. White pixels have been masked out for the computation of the moment maps.}
\label{fig:co_mom}
\end{figure}

In the CO($J=3-2$) low angular resolution map we also detected compact emission coincident with the continuum
emission described in Sect.~\ref{sec:cont}. 
%To highlight this compact emission, we have produced a molecular line emission map using only data from the baselines longer than $\sim 180$~m. In this way, all the CO emission on angular scales larger than about $3''$, or $\sim$ 400~AU at the Ophiuchus distance, is filtered out, and only emission from compact structures is recovered.  
Left panel of Fig.~\ref{fig:co_mom} shows the map of integrated CO($J=3-2$) emission, obtained by filtering out the extended CO emission as described at the end of Section~\ref{sec:obs}.1. 
Compact emission, that we associate with $\rho$-Oph~102, is clearly detected in several channels. The disk is detected at a $> 10\sigma$ level, with a velocity-integrated total flux of about $530 \pm 45$~mJy$\cdot$km/s. Right panel of Fig.~\ref{fig:co_mom}shows the intensity weighted velocity field of CO molecular gas.
%The map shown in Figure~\ref{fig:co_map} also shows the detection of CO emission in correspondence with the location of $\rho$ Oph 102 at the center of the map. We associate this emission to the disk surrounding $\rho$ Oph 102, as done for the dust continuum emission presented in the last Section. In order to better isolate the compact emission of the disk from the more diffuse emission of the molecular cloud visible in Figure~\ref{fig:co_map}, we imaged the CO emission by considering only projected baselines larger than 70 k$\lambda$. In this way, all the CO emission on angular scales larger than about $3''$, or $\sim$ 400~AU at the Ophiuchus distance, is filtered out, and only emission from compact structures is recovered. 
%The only detected emission obtained in this way comes from the $\rho$ Oph 102 disk. In Figure~\ref{fig:co_mom} we present the moment 0 and 1 maps of the the $\rho$ Oph 102 disk in CO($J=3-2$).
%In the moment 0 map the disk is detected at a $> 10\sigma$ level, with a velocity-integrated total flux of about $560 \pm 52$~mJy$\cdot$km/s. The moment 1 map probes the radial velocity field of the source and suggests the 
We tentatively find the presence of a velocity gradient along the disk spanning a range of radial velocities of $\approx 1$~km/s. This is consistent with gas in Keplerian rotation in an inclined disk orbiting at distances $\simgreat~10$~AU from a 0.06~$M_{\odot}$ BD.  
However, the angular ($\sim 0.5''$) and velocity ($\sim 0.4$~km/s) resolutions of our observations do not allow a full characterization of the rotation curve of the disk and non-Keplerian rotation curves are still possible. 

\section{Discussion}
\label{sec:discussion}

%Modelling of the dust thermal emission of protoplanetary disks in the sub-mm continuum is frequently used to investigate dust properties such as mass and physical size of the emitting grains.
To derive the disk mass and physical properties of the grains we model the mm continuum emission using a
two-layer (surface$+$midplane) model of a flared disk heated by the radiation of the central BD \citep{Chiang:1997,Dullemond:2001}.

\begin{figure}
%\plotone{band3}
\epsscale{1.0}
\plotone{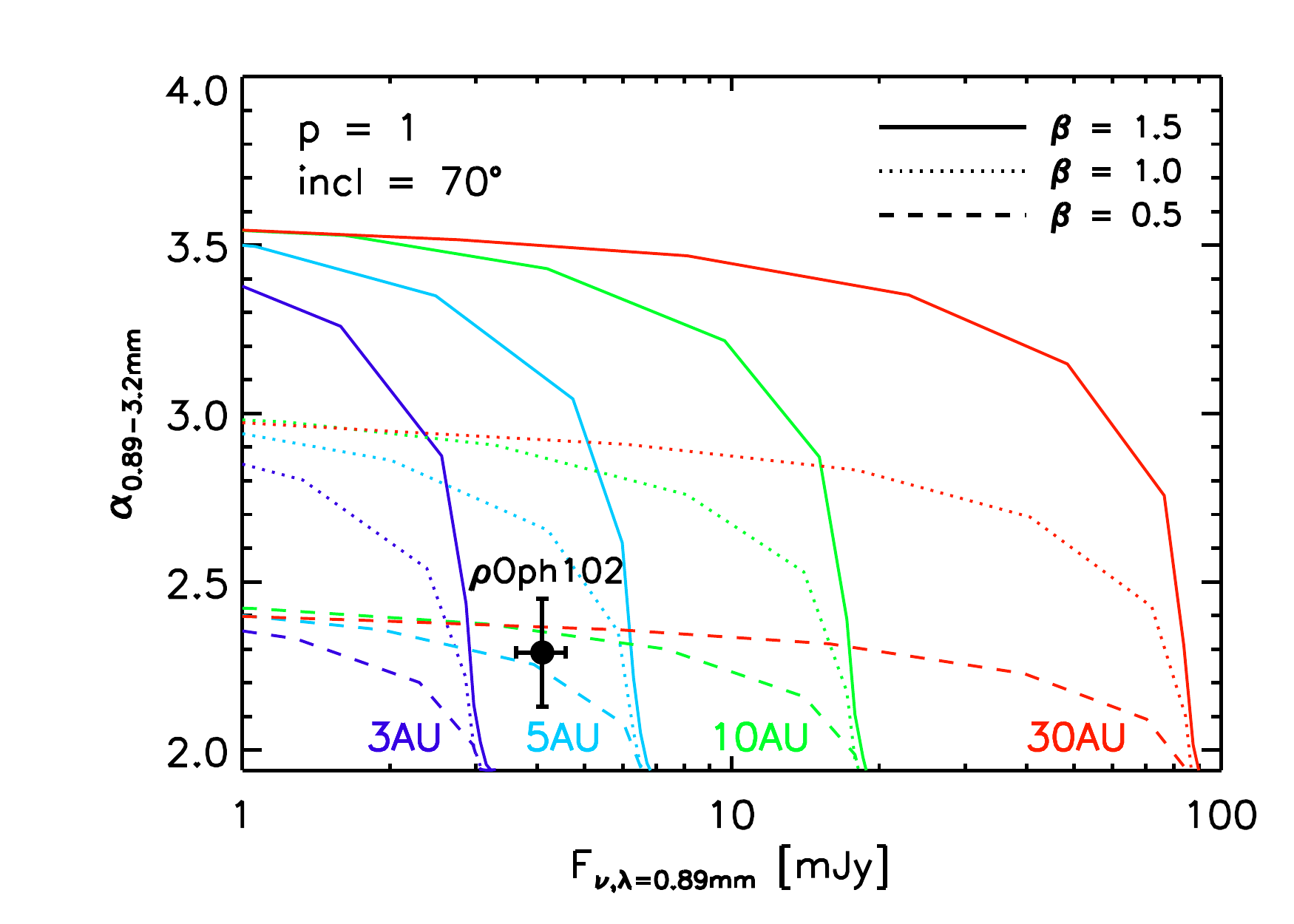}
\caption{Sub-millimeter flux density vs spectral index for disk models for $\rho$-Oph~102. The black dot shows our ALMA data for $\rho$-Oph~102.
Each line represents the prediction of disk models with the same disk outer radius and dust opacity spectral index $\beta$, but increasing disk mass from left to right. The values considered for the disk radii and $\beta$ are indicated. Each model was computed by assuming a radial profile of the surface density $\Sigma \propto r^{-p}$, with $p = 1$, and a disk inclination of 70 degrees. For the sub-stellar properties of $\rho$-Oph~102, we took a mass $M_{\star} = 0.06~M_{\odot}$, luminosity $L_{\star} = 0.08~L_{\odot}$, effective temperature $T_{\rm{eff}} = 2700~$K \citep[from][]{Natta:2002}. For the computation of the dust opacities we considered porous spherical grains made of astronomical silicates, carbonaceous materials and water ice \citep[optical constants for individual components from][respectively]{Weingartner:2001,Zubko:1996,Warren:1984} and adopted a simplified version of the fractional abundances used by \citet[][]{Pollack:1994}, as done in \citet[][]{Ricci:2010a,Ricci:2010b}. Grain sizes are distributed as a power-law $n(a) \propto a^{-q}$ with $q=3.0$ and between a minimum grain size of 0.1 $\mu$m and a maximum grain size which determines the value of $\beta$~\citep[for more details, see][]{Ricci:2010a}.}
\label{fig:models}
\end{figure}

The adopted sub-stellar photospheric properties of  $\rho$-Oph~102  and the dust model are described in the caption of Figure~\ref{fig:models}. Since our observations did not allow us to spatially resolve the dust emission of the disk, we could not strongly constrain the disk radius and inclination. Our analysis of the unresolved disk photometry follows the procedure outlined in \citet{Testi:2001}, adapted for the case of young BDs.  For the disk surface density we adopted a radial profile $\Sigma \propto r^{-p}$ with $p=1$, which is the typical value found for T Tauri disks in the sub-mm~\citep[e.g.][]{Isella:2009,Andrews:2010,Guilloteau:2011}. 
A large value of the disk inclination is suggested by the detection of an optical jet \citep[][]{Whelan:2005} and by the nearly symmetric morphology of the bipolar outflow in molecular CO \citep[][]{Phan-Bao:2008}. At the same time, the visual extinction of the $\rho$-Oph~102 photosphere, i.e. $A_{V} \approx 3$ \citep[][]{Natta:2002}
%, or $A_{H} \approx 0.6$ using the \citet[][]{Cardelli:1989} extinction law with $R_{V} = 3.1$. 
poses an upper limit of $\sim 80$~degrees for the disk inclination, as a more inclined disk would absorb the (sub)-stellar radiation much more efficiently \citep[see][]{Skemer:2011}. We assume in the following an inclination of 70 degrees; however, our results  do not depend critically on this choice. %For the disk radii we considered values of 3, 5, 10, 30~AU, consistently with the upper limit derived in Section~\ref{sec:cont}.

\subsection{Disk mass and size}

The main results of our modeling can be derived from Figure~\ref{fig:models}, which shows the ALMA data overlaid with the predictions from the disk models presented above.
%Following \citet{NattaTesti:2001}, our models include only the heating of the disk from the central photosphere, with this assumption, disks with radii $\simless~3$~AU can not reproduce the observed sub-mm fluxes of the $\rho$-Oph~102 disk. 
%Additional sources of heating for the disk could be external irradiation, which seems unlikely in the environment of our target, or internal due to viscosity. 
%$\rho$-Oph~102 is known to be an accreting object, but the estimated accretion rate is not high enough to imply a significant viscous heating of the disk \citep[][]{Natta:2004,Natta:2006}. 
Disk models with radii larger than $\simgreat~5$~AU can reproduce the data only if the spectral index of the dust opacity $\beta$ ($\kappa_{\nu} \propto \nu^{\beta}$) is smaller than 1. This is because for these disks the dust emission is predominantly coming from the \textit{optically thin} outer regions of the disk. In this case, the spectral index of the SED reflects the dust opacity spectral index, and a relatively low value of $\beta$ is needed to explain the measured spectral index $\alpha_{\rm{0.89-3.2mm}}$. 
Models with disk radii $< 5$~AU reproduce the observed continuum flux and spectral index with a broad range of $\beta-$values. This is because very small disks have to be relatively dense to explain the observed sub-mm fluxes: the dust emission in these models becomes mostly optically thick and, in this regime, it does not depend on the dust opacity law.  \citep[cf. the discussion in][]{Testi:2001}. 
%Data are not consistent with disks smaller than $\sim 3$ AU in radius.

Using the CO($J=3-2$) detection we can test the hypothesis of a small, optically thick disk. The mean brightness temperature of the CO($J=3-2$) in our $\sim 0.5^{\prime\prime}$ FWHM beam is $\sim$8.3~K. As the gas in the disk is primarily heated by radiation from the central BD, its temperature cannot be very high, i.e. $\simless$ 30-40~K, which in turns means that the beam filling factor of the disk cannot be very small. Under the assumptions of optically thick gas emission and face-on geometry, this argument sets a lower-limit of about 15~AU for the disk radius. For example, if the disk radius was 5~AU the beam filling factor would be $<$ 2.3\%, the exact value depending on the disk inclination. This would correspond to a gas temperature $>$ 400~K, which is much higher than expected for a BD disk.

The $\beta$-values constrained by our analysis for these larger disks, i.e. $\beta \approx 0.4 - 0.6$, correspond to dust opacities values of $\kappa^{\rm{dust}}_{\nu,\lambda=0.89\rm{mm}}= 1.5 - 4.5$~cm$^2/$g \citep[see also][]{Ricci:2010a}. Given the measured flux densities, we derive estimates for the dust mass of $\sim 2 - 6\cdot 10^{-6}~M_{\odot}$, or $\sim 2 - 6\cdot 10^{-4}~M_{\odot}$ 
assuming an ISM-like gas-to-dust mass ratio of 100.
%, and of $\sim 6 - 8\cdot 10^{-5}~M_{\odot}$ for the smaller ($3-5$~AU) more optically thick disks with $\beta = 1.5$. 
%Note that for the optically thick disks we  obtain an upper limit for the disk mass. 
%This is because the spectral index measured for the disk is not consistent with the one of an entirely optically thick disk, i.e. 2 for emission in Rayleigh-Jeans regime, or lower than 2 for emission departing from the Rayleigh-Jeans regime.
These values correspond to a disk which contains $\sim$0.3--1\%\ of the mass of the BD$+$disk system and are consistent with the distribution of values for more massive T Tauri systems \citep[][]{WilliamsCieza:2011}.

\subsection{Constraints on the physics of dust evolution in gas-rich disks}

The $\beta$-values needed for disks larger than 5~AU are significantly lower than the values constrained for the ISM \citep[$\approx 1.6-1.8$, see][]{Draine:2003} and are instead consistent with the values derived for disks around PMS stars \cite[$\le 1$, e.g.][]{Natta:2007,Ricci:2010a}. So far, the only plausible hypothesis proposed to explain these values of $\beta$ is that most of the thermal emission observed in the sub-mm is coming from mm-sized grains, or larger, from the disk outer regions \citep[][]{Draine:2006, Ricci:2012}. %These grains are significantly larger than those found in the ISM, and for this reason observations of disks are useful tests 

State-of-the-art models of dust evolution in disks can calculate the size-dependent evolution of solids accounting for a variety of different mechanisms, e.g. coagulation, fragmentation, radial migration \citep[][]{Brauer:2008,Birnstiel:2010a}. Sub-mm observations providing information on the grain size distribution of dust in disks around PMS stars have been used to test the predictions from these models \citep[][]{Birnstiel:2010b,Pinilla:2012}. In these models, the efficiency for a disk to grow solids critically depends on its physical conditions. If compared with more massive disks around PMS stars, these models predict that at the conditions of BD disks (e.g. lower densities, higher relative velocities due to radial drift) the growth to mm-grains should be much less efficient. 
Furthermore, in BD disks inward radial drift of $\sim$mm-grains is expected to be faster than in T Tauri disks by a factor of a few.
Therefore, the discovery of mm-sized grains in the outer regions of BD disks, as suggested in the case of $\rho$-Oph~102, severely challenges models of early evolution of solids.
Possible ways to explain the presence of mm-sized grains in the outer regions of young BD disks are discussed in Pinilla et al. (2012, in prep.). These include very low levels of turbulence in the disk to decrease the collisional speed between solids, and require the action of some physical mechanisms which can efficiently halt the inward radial migration of these particles, e.g. gravitational clumping favored by the magnetic field in the disk \citep[][]{Johansen:2007}. 

% A general prediction of these models is that the maximum grain size which can be reached with growth by sticking should scale linearly with the gas density at a given location in the disk. At the lower

% One expectation
%of these models is that to grow mm-size grains in the outer regions of disks around brown dwarfs should be much more difficult than around solar-mass stars, assuming that transitioning to lower and lower system mass there is no significant discontinuity in the ratio of M$_{disk}$/M$_\star$, the disk geometry, the turbulence in the disk and the dust sticking and fragmentation parameters. 

%Therefore, the largest grains present in the outer regions of disks around brown dwarfs should be significantly smaller than those found in the denser disks  around PMS stars.

\section{Summary}

We presented ALMA observations of the young Brown Dwarf $\rho$-Oph~102 at about 0.89~mm and 3.2~mm. 
We reported the detection of the disk surrounding $\rho$-Oph~102 in both the dust continuum bands and in CO($J = 3 - 2$).   
Dust continuum emission is spatially unresolved, and this provides an upper limit on the radius of the dusty disk of $\sim 40$~AU.
We derived a disk mass of $\sim 0.3-1$\%\ of the mass of the Brown Dwarf, similar to the typical values found for disks around PMS stars.
The detection of CO molecular gas emission at the location of the Brown Dwarf indicates a gas-rich disk as typically found for disks surrounding more massive young stars. 
%From the CO emission we derived a lower limit on the radius of the gaseous disk of $\sim 15$~AU.
We measured a spectral index of the continuum emission $\alpha_{\rm{0.89-3.2mm}} = 2.29 \pm 0.16$, which is significantly lower than the value found in the ISM. By modeling the dust emission and from the properties of the CO molecular gas emission we constrained the variation of the dust opacity with frequency, which is a diagnostic of grain growth in the disk. Our result indicates that mm-grains are found not only in the outer regions of T Tauri disks, but even in the low density environments of Brown Dwarf disks, and this challenges our current understanding of dust evolution in gas-rich disks.

\acknowledgments We thank P. Pinilla, T. Birnstiel and J. Carpenter for helpful comments, E. van Kampen and ESO ARC for technical support. L. R. acknowledges support from the ESO Scientific Visitor Programme.
I. G. is supported by the Spanish MINECO grant AYA2011-30228-C03-01 (co-funded with FEDER fund). 
This paper makes use of the following ALMA data: ADS/JAO.ALMA\#2011.0.00259.S. ALMA is a partnership of ESO (representing its member states), NSF (USA) and NINS (Japan), together with NRC (Canada) and NSC and ASIAA (Taiwan), in cooperation with the Republic of Chile. The Joint ALMA Observatory is operated by ESO, AUI/NRAO and NAOJ. The National Radio Astronomy Observatory is a facility of the National Science Foundation operated under cooperative agreement by Associated Universities, Inc.

%\begin{figure}
%%\plotone{band3}
%\epsscale{1.0}
%\plotone{co_map}
%\caption{Integrated CO($J=3-2$) map of the region surrounding $\rho$ Oph 102. The map was obtained by integrating the CO($J=3-2$) emission pixel by pixel. For the imaging, we adopted Briggs weighting with robust parameter $= 2$ (natural weighting) and a Gaussian 1 arcsec uv-taper on the outer baselines to filter out emission from small scales ($< 1''$) and highlight the diffuse emission from the region. The field is centered at the location of $\rho$ Oph 102 and shows the entire field of view of our ALMA observations. The size of the synthesized beam is shown in magenta in the lower left corner (FWHM $= 1.3'' \times 1.2''$, PA $= 94$~deg).}
%\label{fig:co_map}
%\end{figure}

\end{document}